\documentclass[prd,nofootinbib,preprint,superscriptaddress]{revtex4}

\usepackage{amsmath, amssymb, amsthm, graphicx, epsfig, fancyhdr,epsfig, slashed}

\usepackage{tikzsymbols}
\usepackage{natbib}
\usepackage{float}
\usepackage{xcolor}
\usepackage{tikz,xcolor,hyperref}
\usepackage{mathtools}

\definecolor{lime}{HTML}{A6CE39}
\DeclareRobustCommand{\orcidicon}{
	\begin{tikzpicture}
	\draw[lime, fill=lime] (0,0) 
	circle [radius=0.2] 
	node[white] {{\fontfamily{qag}\selectfont \tiny ID}};
	\draw[white, fill=white] (-0.0625,0.095) 
	circle [radius=0.007];
	\end{tikzpicture}
	\hspace{-2mm}
}

\foreach \x in {A, ..., Z}{\expandafter\xdef\csname orcid\x\endcsname{\noexpand\href{https://orcid.org/\csname orcidauthor\x\endcsname}
			{\noexpand\orcidicon}}
}

\newcommand{\be}{\begin{equation}}
\newcommand{\ee}{\end{equation}}
\newcommand{\bea}{\begin{eqnarray}}
\newcommand{\eea}{\end{eqnarray}}

\newcommand{\pfrac}[2]{\left(\frac{#1}{#2}\right)}
\newcommand{\GeV}{{\rm\,GeV}}
\newcommand{\sn}{\mathop{\rm sn}\nolimits}

\newcommand{\bbbone}{\hbox{\rm 1\kern-3pt l}}
\newcommand{\slp}{p\kern-5pt/}

\newcommand{\tr}{\mathop{\rm tr}\nolimits}

\begin{document}

\title{Confinement in QCD and generic Yang-Mills theories
  with matter representations}

\author{Marco Frasca\orcidA{}}
\email{marcofrasca@mclink.it}
\affiliation{Rome, Italy}

\author{Anish Ghoshal\orcidB{}}
\email{anish.ghoshal@fuw.edu.pl}
\affiliation{Institute of Theoretical Physics, Faculty of Physics,
 University of Warsaw, ul.\ Pasteura 5, 02-093 Warsaw, Poland}

\author{Stefan Groote\orcidC{}}
\email{stefan.groote@ut.ee}
\affiliation{F\"u\"usika Instituut, Tartu Ulikool,
  W.~Ostwaldi 1, EE-50411 Tartu, Estonia}

\begin{abstract}
\textit{We derive the low-energy limit of quantum chromodynamics (QCD) and
provide evidence that in the 't~Hooft limit, i.e.\ for a very large number
of colors and increasing 't~Hooft coupling, quark confinement is recovered.
The low energy limit of the theory turns out to be a non-local
Nambu--Jona-Lasinio (NJL) model. The effect of non-locality, arising from a
gluon propagator that fits quite well to the profile of an instanton liquid,
is to produce a phase transition from a chiral condensate to an instanton
liquid, as the coupling increases with lower momentum. This phase transition
suffices to move the poles of the quark propagator to the complex plane. As a
consequence, free quarks are no longer physical states in the spectrum of the
theory.}
\end{abstract}

\maketitle

\section{Introduction}

One of the most important open problems in physics is the question of quark
confinement. Quarks are never seen free but only in bound states (cf.\
Ref.~\cite{Greensite:2011zz,Kogut:2004su} and references therein). Several
mechanisms have been proposed but none of them has ever been derived fully
analytically from theory, i.e.\ quantum chromodynamics (QCD). An exception can
be found in Kenneth G.\ Wilson's semianalytic approach to QCD regularised on
the lattice for which an area law for confinement at strong couplings has been
demonstrated~\cite{Wilson:1974sk}. Some criteria have been obtained for
confinement in four dimensions. Kugo and Ojima were able to obtain a
well-known criterion starting from a reformulation of BRST
invariance~\cite{Kugo:1979gm,Kugo:1977zq}. Similarly, Nishijima and his
collaborators pointed out some constraints to grant
confinement~\cite{Nishijima:1993fq,Nishijima:1995ie,Chaichian:2000sf,%
Chaichian:2005vt,Nishijima:2007ry}. A proof of confinement exists in
supersymmetric models where a condensate of monopoles like in a type~II
superconductors is seen~\cite{Seiberg:1994aj,Seiberg:1994rs}. Indeed, the
exact $\beta$ function for the Yang-Mills theory is known for some
supersymmetric and the non-supersymmetric models~\cite{Novikov:1983uc,%
Shifman:1986zi,Ryttov:2007cx,Chaichian:2018cyv}. Lattice simulations have
calculated the Yang-Mills theory beta function by using the RG evolution (in
several schemes) in order to connect the regularisation scale with the
infrared behavior, cf.\ e.g. Ref.~\cite{Hasenfratz:2023bok}. By tuning
empirical parameters, the beta functions of the models can be made consistent
with the lattice results. Different confinement criteria and their overlapping
regions are presented in Ref.~\cite{Chaichian:1999is}.

Due to the discovery of Gribov copies~\cite{Gribov:1977wm} and their possible
handling as proposed by Zwanziger~\cite{Zwanziger:1989mf}, studies on
confinement in Landau gauge seemed to indicate a gluon propagator running to
zero in the infrared while the ghost propagator had to run to infinity faster
than in the free case. This qualitative picture is essential to have an idea
of the potential between quarks. Measures of the gluon and ghost
propagators~\cite{Bogolubsky:2007ud,Cucchieri:2007md,Oliveira:2007px} and the
spectrum~\cite{Lucini:2004my,Chen:2005mg} on the lattice have shown that in a
non-Abelian gauge theory without fermions a mass gap appears, in evident
contrast with the scenario devised by Gribov and Zwanziger that in the
original formulation is not able to accommodate this mass gap. As shown in
several theoretical works, the behavior seen on the lattice should be
expected~\cite{Cornwall:1981zr,Cornwall:2010bk,Dudal:2008sp,Frasca:2007uz,%
Frasca:2009yp,Frasca:2015yva}. These works provide closed form formulas for
the gluon propagator with a number of fitting parameters. Indeed, a closed
analytical formula for the gluon propagator is an important element to obtain
the low-energy behavior of QCD in a manageable effective theory to prove
confinement. For the same aim, the behavior of the running coupling in the
infared limit is essential~\cite{Bogolubsky:2009dc,Duarte:2016iko} (see also
the review~\cite{Deur:2016tte}). An instanton liquid picture seems to play a
relevant role~\cite{Schafer:1996wv,Boucaud:2002fx}. Confinement in its
simplest form can be seen as the combined effect of a potential obtained from
the Wilson loop of a Yang-Mills theory without fermions and the running
coupling yielding a linearly increasing potential in agreement with lattice
data~\cite{Deur:2016bwq}. Note that in spacetime 2+1 dimensions, the theory is
only marginally confining, as there is no running coupling and the potential
increases only logarithmically. Still, also in this lower dimension
confinement is granted~\cite{Frasca:2016sky}. 

An essential part in our understanding of confinement in QCD is strongly
linked to a proper derivation of the low-energy limit of the theory. In this
direction, a couple of seminal papers were written by Gerard 't~Hooft for 1+1
spacetime dimensions~\cite{tHooft:1973alw,tHooft:1974pnl}. Two important
results were obtained by 't~Hooft in these papers: 1) A good understanding of
the low-energy behavior of the theory could be obtained in principle by
considering the limit of the number of colors $N_c$ running to infinity and
keeping the product $N_cg^2$ constant, with $\alpha_s=g^2/(4\pi)$ the strong
coupling. In this limit the coupling goes to zero, faciliating a perturbative
approach. 2) In 1+1 spacetime dimensions the theory can be solved and provides
the meson spectrum of the theory. One of the conclusions for 3+1 spacetime
dimensions was that a discrete spectrum of the Hamiltonian can grant the
appearance of condensates providing the right set-up for quark confinement
through a string model~\cite{tHooft:1973alw,Brodsky:2012ku}. Indeed, in
Refs.~\cite{Frasca:2016sky,Frasca:2017slg} a discrete spectrum was obtained
for a Yang-Mills theory without quarks, confirming lattice results. Without
providing explicitly the spectrum, this was also proved
mathematically~\cite{Dynin:2017}.

In a series of works~\cite{Frasca:2015yva,Frasca:2021zyn,Frasca:2021yuu}, it
was recently proved that in the 't~Hooft limit a of large number of colors the
low energy limit of QCD is given by a non-local Nambu--Jona-Lasinio (NJL)
model~\cite{Nambu:1961tp,Nambu:1961fr,Klevansky:1992qe,GomezDumm:2006vz,%
Hell:2008cc}. The local version of the NJL-model, as initially conceived in
Refs.~\cite{Nambu:1961tp,Nambu:1961fr}, does not confine. For bounded states
obtained after bosonization~\cite{Ebert:1997fc}, there is a threshold for the
decay into quark and antiquark as free states that have never been observed.
Non-locality can help to remove such a problem~\cite{Bowler:1994ir}. The aim
of this paper is to provide evidence that the non-local NJL-model derived from
QCD in Refs.~\cite{Frasca:2015yva,Frasca:2021zyn,Frasca:2021yuu} is indeed
confining with the principles given in Refs.~\cite{Bowler:1994ir,%
Roberts:1994dr}. 

This paper is phenomenological in nature, and some relevant approximations are
involved to solve the QCD equations in the low-energy limit. The most relevant
of these is the 't Hooft limit $N_c$ with the 't Hooft coupling $\lambda=N_cg^2$
kept finite but large. Accordingly, we expand in $1/\lambda$ and terminate the
expansion at leading order, neglecting higher-order correlations between
fermionic degrees of freedom caused by the gluonic field.

The paper is structured as follows. In Sec.~\ref{sec2}, we derive the
NJL-model from QCD and apply the bosonization and the mean field approximation,
leading to the gap equation. In Sec.~\ref{sec3}, we present the proof of quark
confinement for the low-energy limit of QCD based on the gap equation we
obtained previously. In Sec.~\ref{conc} we give our conclusions.

\medskip


\section{Low-energy limit of QCD and NJL Model\label{sec2}}

We consider the QCD lagrangian
\begin{equation}
{\cal L}_{QCD}=\sum_i\bar q_i(i\gamma^\mu D_\mu+m)q_i
  -\frac14F^{\mu\nu}_aF_{\mu\nu}^a
  -\frac1{2\xi}(\partial_\mu A^\mu_a)(\partial_\nu A^\nu_a)
\end{equation}
where $D_\mu=\partial_\mu+igT_aA_\mu^a$ is the covariant derivative and
$F_{\mu\nu}^a$ are the field strenght tensor components. These can be obtained
by $igT_aF_{\mu\nu}^a=[D_\mu,D_\nu]$. The sum over $i$ is quite generic as it
implies the sum over quark flavors and colors. Our Minkowskian metric reads
$g^{\mu\nu}=\mbox{\rm diag}(1,-1,-1,-1)$.

From the Euler--Lagrange equations, we obtain
\begin{eqnarray}
0&=&\frac{\partial{\cal L}_{QCD}}{\partial A_\nu^a}
  -\partial_\mu\frac{\partial{\cal L}_{QCD}}{\partial(\partial_\mu A_\nu^a)}
  \nonumber\\
  &=&\partial_\mu(\partial^\mu A^\nu_a-\partial^\nu A^\mu_a)
  +\frac1\xi\partial^\nu(\partial_\mu A^\mu_a)
  +gf_{abc}\partial_\mu(A^\mu_bA^\nu_c)\strut\nonumber\\&&\strut
  +gf_{abc}(\partial^\mu A^\nu_b-\partial^\nu A^\mu_b)A_\mu^c
  +g^2f_{abc}f_{cde}A^\mu_bA^\nu_dA_\mu^e-g\sum_i\bar q_i\gamma^\nu T_aq_i,
  \nonumber\\
0&=&\frac{\partial{\cal L}_{QCD}}{\partial\bar q_i}
  -\partial_\mu\frac{{\cal L}_{QCD}}{\partial(\partial\bar q_i)}
  \ =\ (i\gamma^\mu D_\mu+m)q_i.
\end{eqnarray}
From the equations of motion we can derive, in principle, the full hierarchy
of Dyson--Schwinger equations. We solve this hierarchy by a method proposed by
Bender, Milton and Savage~\cite{Bender:1999ek}, recently exploited in
Refs.~\cite{Frasca:2013kka,Frasca:2013tma,Frasca:2015yva,Frasca:2017slg,%
Frasca:2019ysi}. Note that if a source term is added to the Lagrangian
describing the vacuum expectation values, translational invariance is broken,
as it is expressed by the separate arguments in the Green functions. In this
case, the vacuum expectation values of products of field operators expressing
those Green functions are nonvanishing even in the case of a single field
operator. This gives sense to starting to solve the tower of Dyson--Schwinger
equations with just this one-point Green function. In the end, setting the
source term to zero restores the observable physical picture. This procedure
is quite similar to the approach provided by the generating functional.

Accordingly, to the Lagrangian we add source terms like $A_\mu^aJ^\mu_a$,
$\bar q_i\eta_i$ and $\bar\eta_iq_i$. For the sake of simplicity, we omit
details on BRST ghosts. After this addition we can evaluate the functional
derivatives with respect to these sources. Such a procedure yields the
Dyson--Schwinger equations~\cite{Frasca:2019ysi}
\begin{eqnarray}
\lefteqn{\partial^2A_\nu^{1a}(x)+gf_{abc}
  \left(\partial^\mu A_{\mu\nu}^{2bc}(x,x)+\partial^\mu A_\mu^{1b}(x)
  A_\nu^{1c}(x)-\partial^\nu A_{\mu\nu}^{2bc}(x,x)
  -\partial_\nu A_\mu^{1b}(x)A_c^{1\mu}(x)\right)\strut}\nonumber\\&&\strut
  +gf_{abc}\partial^\mu A_{\mu\nu}^{2bc}(x,x)
  +gf_{abc}\partial^\mu(A_\mu^{1b}(x)A_\nu^{1c}(x))	
  +g^2f_{abc}f_{cde}\Big(g^{\mu\rho}A_{\mu\nu\rho}^{3bde}(x,x,x)
  \strut\nonumber\\&&\strut
  +A_{\mu\nu}^{2bd}(x,x)A_e^{1\mu}(x)+A_{\nu\rho}^{2eb}(x,x)A_d^{1\rho}(x)
  +A_{\mu\nu}^{2de}(x,x)A_b^{1\mu}(x)+A_b^{1\mu}(x)A_\mu^{1d}(x)A_\nu^{1e}(x)
  \Big)\nonumber\\
  &=&g\sum_i\gamma_\nu T^aq_{ii}^2(x,x)+g\sum_i\bar q_i^1(x)\gamma_\nu
  T^a q_i^1(x),\nonumber\\
\lefteqn{(i\slashed\partial-m_q)q_i^1(x)
  +g\gamma^\mu A_\mu^{1a}(x)T_aq_i^1(x)\ =\ 0,}
\end{eqnarray}
where we have introduced the one-, two- and three-point functions as 
$A_\mu^{1a}(x)=\langle A_\mu^a(x)\rangle$,
$A_{\mu\nu}^{2ab}(x,y)=\langle A_\mu^a(x)A_\nu^b(y)\rangle$ and
$A_{\mu\nu\rho}^{3abc}(x,y,z)=\langle A_\mu^a(x)A_\nu^b(x)A_\rho^c(x)\rangle$
for the gauge fields, and $q_i^1(x)=\langle q_i(x)\rangle$ and
$q_{ij}^2(x,y)=\langle q_i(x)q_j(y)\rangle$ for the quark fields. We can use
the exact solutions already provided in Ref.~\cite{Frasca:2015yva} to write
\begin{equation}
\label{eq:map}
A_\nu^{1a}(x)=\eta_\nu^a\phi(x),\qquad
A_{\mu\nu}^{2ab}(x,y)=\left(g_{\mu\nu}
  -\frac{\partial_\mu\partial_\nu}{\partial^2}\right)\delta^{ab}\Delta(x-y),
\end{equation}
where $\eta_\mu^a$ are the coefficients of the polarization vector with
$\eta_\mu^a\eta^\mu_b=\delta_{ab}$, $\phi(x)$ is a scalar field and
$\Delta(x-y)$ is the propagator of the scalar field. The ansatz we can afford
so far, namely the gluon field as a constant polarization vector times a
scalar function, is based on Refs.~\cite{Frasca:2007uz,Frasca:2009yp}\footnote{
The mapping between a quartic scalar field and the Yang-Mills theory has been a
matter of discussion with the mathematician Terence Tao (Fields medallist) who
accepted the proof given in Ref.~\cite{Frasca:2009yp} showing that such a
mapping is exact in the Landau gauge but just asymptotic for other gauge
choices for the coupling running to infinity.}, and can help us to reach up to
essential statements on the confinement. One obtains
\begin{eqnarray}\label{Dirac}
\eta_\nu^a\partial^2\phi(x)+2N_cg^2\Delta(0)\eta_\nu^a\phi(x)
  +N_cg^2\eta_\nu^a\phi^3(x)&=&g\sum_i\gamma_\nu T_aq_{ii}^2(x,x)
  +g\sum_i\bar q_i^1(x)\gamma_\nu T^aq_i^1(x),\nonumber\\
(i\gamma^\mu\partial_\mu-m_q)q_i^1(x)+g\gamma^\mu\eta_\mu^aT_a\phi(x)q_i^1(x)
  &=&0.
\end{eqnarray}
Using the properties of the symbols $\eta$, one has
$\eta_\mu^a\eta^\mu_a=N_c^2-1$ and $\sum_iq_{ii}^2(x,x)=N_cN_fS(0)$.
Accordingly, the first differential equation~(\ref{Dirac}) takes the form
\begin{equation}\label{first}
\partial^2\phi(x)+2N_cg^2\Delta(0)\phi(x)+N_cg^2\phi^3(x)
  =\frac{g}{N_c^2-1}\left[N_cN_f\gamma^\nu\eta_\nu^aT_aS(0)
  +\sum_i\bar q_i^1(x)\gamma^\nu\eta_\nu^aT_aq_i^1(x)\right].\kern-5pt
\end{equation}

At this point we can consider the 't~Hooft limit $N_c\to\infty$ with
$\lambda:=N_cg^2\gg 1$ being finite but large. Next, we can rescale the space
variable as $x\rightarrow\sqrt{N_cg^2}x$ and look for a perturbative series
$\phi(x)=\phi_0(x)+(N_cg^2)^{-1}\phi_1(x)+O((N_cg^2)^{-2})$ in the 't Hooft
coupling, yielding at leading order (after reverting the rescaling)
\begin{equation}
\partial^2\phi_0(x)+2\lambda\Delta(0)\phi_0(x)+\lambda\phi_0^3(x)=0,
\end{equation}
while the next-to-leading order yields
\begin{eqnarray}\label{NLO}
\lefteqn{\partial^2\phi_1(x)+2\lambda\Delta(0)\phi_1(x)
  +3\lambda\phi_0^2(x)\phi_1(x)\ =}\nonumber\\
  &=&\frac{g}{N_c^2-1}\left[N_cN_f\gamma^\nu\eta_\nu^aT_aS(0)
  +\sum_i\bar q_i^1(x)\gamma^\nu\eta_\nu^aT_aq_i^1(x)\right].
\end{eqnarray}
Truncating the series expansion in inverse powers of $N_cg^2$ at the first
order, one ends up with a NJL model. Higher orders would generate interactions
with more than four fermions involved. An attempt in this direction was
presented in Ref.~\cite{Frasca:2016lit} where the next-to-leading order terms
turn out to depend on products of the gluon Green functions and higher powers
of pairs of fermionic fields.

\subsection{Zeroth order solution and Green function}

As a constant, $m^2=2\lambda\Delta(0)$ can be considered as the mass squared.
Even though the leading order equation
$\partial^2\phi_0(x)+m^2\phi_0(x)+\lambda\phi_0^3(x)=0$ is nonlinear, we have
found a solution expressed by Jacobi's elliptic function~\cite{Frasca:2009bc},
\begin{equation}
\phi_0(x)=\mu\sn\left(k\cdot x+\theta\,|\,\kappa\right)
\end{equation}
with
\begin{equation}
k^2=\frac12\lambda\mu^2+m^2\quad\mbox{and}\quad
\kappa=\frac{m^2-k^2}{k^2},
\end{equation}
where $\mu$ and $\theta$ are integration constants. $\sn(z|\kappa)$ is
Jacobi's elliptic function of the first kind. Inserting this solution into the
Green's equation
\begin{equation}
\left(\partial^2+m^2+3\lambda\phi_0^2(x)\right)\Delta(x-y)=\delta^4(x-y)
\end{equation}
for the two-point function obtained as the next element of the tower of
Dyson--Schwinger equations, this equation can be solved in momentum space by
\begin{equation}\label{Delta0mom}
\tilde\Delta(p)=
  -\sum_{n=0}^\infty\frac{B_n(\kappa)}{p^2-m_n^2+i\epsilon},\qquad
  B_n(\kappa)=\frac{(2n+1)^2\pi^3}{2(1-\kappa)K(\kappa)^3\sqrt\kappa}
  \frac{(-1)^ne^{-(n+1/2)\varphi(\kappa)}}{1-e^{-(2n+1)\varphi(\kappa)}},
\end{equation}
where
\begin{equation}
\varphi(\kappa)=\frac{K^*(\kappa)}{K(\kappa)}\pi,\qquad K^*(z)=K(1-z).
\end{equation}
The corresponding mass spectrum reads
\begin{equation}
\label{eq:spec}
m_n=\frac{(2n+1)\pi}{2K(\kappa)}\sqrt{2k^2}=:(2n+1)m_G(\kappa).
\end{equation}
This procedure ends up with a gap equation by inserting the Fourier transform
of the propagator~(\ref{Delta0mom}) back into $m^2=2\lambda\Delta(0)$, yielding
\begin{equation}\label{m2}
m^2=-2\lambda\int\frac{d^4p}{(2\pi)^4}\sum_{n=0}^\infty
  \frac{B_n(\kappa)}{p^2-(2n+1)^2m_G^2(\kappa)+i\epsilon}.
\end{equation}
It can be shown that by this gap equation the spectrum of the theory without
fermions is correctly given~\cite{Frasca:2017slg} in excellent agreement with
lattice data. Note that the momentum scale $k^2$ of the one-point function
and the momentum scale $p^2$ of the two-point function are independent.
Because of this, in the following section we use an empirical value to fix
$k^2$.

In order to complete this section, we argue that the zeros of the gluon
propagator are genuine glueball colorless states. We start by considering the
correlation function for the scalar glueballs that is given
by~\cite{Narison:2002woh,Narison:2021xhc}
\begin{equation}
{\cal O}(x)=\langle F^{a\mu\nu}(x)F^a_{\mu\nu}(x)
  F^{b\rho\eta}(0)F^b_{\rho\eta}(0)\rangle.
\end{equation}
Using methods explained in Ref.~\cite{Frasca:2015yva}, one can see that
according to Ref.~\cite{Windisch:2012sz} the four-point correlator defining
the correlation function of the glueball can be reduced to convolutions over
one- and two-point functions. As the one-point function has no poles but
zeros, the poles of the glueball four-point correlator are given by the poles
of the two-point correlator. Therefore, these poles represent true colorless
glueball states.

\subsection{First order solution}

By convoluting the propagator $\Delta$ with the right hand side of
Eq.~(\ref{NLO}) we obtain
\begin{eqnarray}\label{eq:phi_1}
  \phi_1(x)=\frac{g}{N_c^2-1}\int d^4y\Delta(x-y)
  \left[N_cN_f\gamma^\nu\eta_\nu^aT_aS(0)
  +\sum_i{\bar q}_i^1(y)\gamma^\nu\eta_\nu^aT_aq_i^1(y)\right].
\end{eqnarray}
We observe that the first term is just a renormalization of the fermion mass
and can be chosen to be zero via the condition $S(0)=0$. The second term is
the expected NJL interaction in the equation of motion of the quarks.

The solution $\phi(x)$ we obtained above can be inserted into Eq.~(\ref{Dirac}).
In the 't Hooft limit, we note that the term $\phi_0$ is negligibly small
compared to the NJL term $\phi_1$. One can see this by observing that
$\phi_0\sim\lambda^{1/4}$ and $\phi_1\sim\lambda$. Therefore, in the strong
coupling limit $\lambda\gg 1$ the equation for
the one-point function of the quark has just the NJL term. We can write
\begin{equation}
(i\gamma^\mu\partial_\mu-m_q)q_i^1(x)+\frac{g^2}{N_c^2-1}\sum_\eta\int d^4y
  \Delta(x-y)\gamma^\mu\eta_\mu^aT_aq_i^1(x)
  \sum_j\left[{\bar q}_j^1(y)\gamma^\nu\eta_\nu^bT_bq_j^1(y)\right]=0.
\end{equation}
Such an equation can be recognized as the Euler--Lagrange equation for the
one-point function of the quark obtained from a NJL model with a non-local
interaction~\cite{Bowler:1994ir,GomezDumm:2006vz}
\begin{eqnarray} \label{NJLeq1}
\lefteqn{{\cal L}''_{\rm NJL}
  \ =\ \sum_i\bar q_i^1(x)(i\gamma^\mu\partial_\mu-m_q)q_i^1(x)
  \strut}\nonumber\\&&\strut
  +\frac{g^2}{N_c^2-1}\sum_\eta
  \sum_i\left[\bar q_i^1(x)\gamma^\mu\eta_\mu^aT_aq_i^1(x)\right]
  \int d^4y\Delta(x-y)
  \sum_j\left[\bar q_j^1(y)\gamma^\nu\eta_\nu^bT_bq_j^1(y)\right].
\end{eqnarray}
We see that $\sum_\eta\eta_\mu^a\eta_\nu^b=\delta_{ab}g_{\mu\nu}$, where
$\eta$ symbolizes the polarizations. Tracing out the color degrees of freedom
with $\tr(T_aT_a)=N_cC_F$, $C_F=(N_c^2-1)/(2N_c)$, and
\[\sum_i\bar q_i^1(x)\gamma^\mu q_i^1(x)
  =N_c\sum_i\bar\psi_i(x)\gamma^\mu\psi_i(x),\qquad
  \sum_i\bar q_i^1(x)q_i^1(x)=N_c\sum_i\bar\psi_i(x)\psi_i(x),\]
with $\psi_i(x)$ being spinors in Dirac and flavor space only, we are led to
the NJL Lagrangian
\begin{eqnarray}
\lefteqn{{\cal L}'_{\rm NJL}\ =\ \sum_i\bar\psi_i(x)
  (i\gamma^\mu\partial_\mu-m_q)\psi_i(x)\strut}\nonumber\\&&\strut
  +\frac{N_cg^2}2\sum_i\left[\bar\psi_i(x)\gamma^\mu\psi_i(x)\right]
  \int d^4y\Delta(x-y)\sum_j\left[\bar\psi_j(y)\gamma_\mu\psi_j(y)\right].
\end{eqnarray}
After a Fierz rearrangement of the quark fields we obtain
(cf.\ e.g.\ Refs.~\cite{Cahill:1985mh,Roberts:1985ju,Roberts:1986ps,%
Praschifka:1986nf,Hell:2008cc})
\begin{eqnarray}\label{fierz}
\lefteqn{{\cal L}'_{\rm NJL}\ =\ \sum_i\bar\psi_i(x)
  (i\gamma^\mu\partial_\mu-m_q)\psi_i(x)\strut}\nonumber\\&&\strut
  +\frac{N_cg^2}2\int d^4y\Delta(x-y)\sum_{i,j}\bar\psi_i(x)\psi_j(y)
  \bar\psi_j(y)\psi_i(x)\strut\nonumber\\&&\strut
  +\frac{N_cg^2}2\int d^4y\Delta(x-y)\sum_{i,j}\bar\psi_i(x)i\gamma_5\psi_j(y)
  \bar\psi_j(y)i\gamma_5\psi_i(x)\strut\nonumber\\&&\strut
  -\frac{N_cg^2}4\int d^4y\Delta(x-y)\sum_{i,j}\bar\psi_i(x)\gamma^\mu
  \psi_j(y)\bar\psi_j(y)\gamma_\mu \psi_i(x)\strut\nonumber\\&&\strut
  -\frac{N_cg^2}4\int d^4y\Delta(x-y)\sum_{i,j}
  \bar\psi_i(x)\gamma^\mu\gamma_5\psi_j(y)
  \bar\psi_j(y)\gamma_\mu\gamma_5\psi_i(x).
\end{eqnarray}

\subsection{Bosonization}

$\Gamma_\alpha$ are understood as a set of combined Dirac and flavor matrices 
given by $1$, $i\gamma_5$, $\gamma_\mu$ and $\gamma_\mu\gamma_5$ after the
Fierz rearrangement and the flavor matrices $\bbbone$ and
$\frac12\lambda_\alpha$ relating quarks of equal and different flavor $i$ and
$j$ in adjoint representation. For $\Gamma_\alpha$ we have the conjugation rule
$\gamma^0\Gamma_\alpha^\dagger\gamma^0=\Gamma_\alpha$, where $\alpha$ denotes
the components of the adjoint flavor representation. Accordingly, the spinor
$\psi(x)$ spans over all these spaces. As the coefficients of these two
contributions are the same, the sum over these given $1+3=4$ degrees of
freedom can be reinterpreted as a sum over the components of a four-vector.
Next we apply the bosonization procedure shown in Ref.~\cite{Hell:2008cc} by
adding scalar--isoscalar and pseudoscalar--isovector mesonic fields at an
intermediate space-time location $w=(x+y)/2$ as auxiliary fields
$M_\alpha(w)=(\sigma(w);\vec\pi(w))$ coupled to the nonlocal fermionic
currents. After Fierz rearrangement, this sums up to the NJL action
\begin{eqnarray}
\lefteqn{{\cal S}_{\rm NJL}\ =\ -\frac{N_cg^2}{2G^2}\int d^4z\Delta(z)
  \int d^4wM_\alpha^*(w)M^\alpha(w)\strut}\\&&\strut
  +\int d^4x\Bigg[\bar\psi(x)(i\gamma^\mu\partial_\mu-m_q)\psi(x)
  +\frac{N_cg^2}2\int d^4y\Delta(x-y)\bar\psi(x)\Gamma_\alpha\psi(y)
  \bar\psi(y)\Gamma^\alpha\psi(x)\Bigg]\nonumber
\end{eqnarray}
($G=2\int d^4z\Delta(z)=2\tilde\Delta(0)$). By performing a nonlocal
functional shift
\begin{equation}
M_\alpha\pfrac{x+y}2\to M_\alpha\pfrac{x+y}2+G\bar\psi(x)\Gamma_\alpha\psi(y),
\end{equation}
the nonlocal quartic fermionic interaction can be removed. Instead, the
fermion field starts to interact nonlocally with the mesonic fields,
\begin{eqnarray}
{\cal S}_{\rm NJL}&=&-\frac{N_cg^2}{2G^2}\int d^4z\Delta(z)
  \int d^4wM_\alpha^*(w)M^\alpha(w)+\int d^4x\bar\psi(x)
  (i\gamma^\mu\partial_\mu-m_q)\psi(x)\strut\\&&\strut
  -\frac{N_cg^2}{2G}\int d^4x\int d^4y\bar\psi(x)\Delta(x-y)
  \left(M_\alpha\pfrac{x+y}2+M_\alpha^*\pfrac{x+y}2\right)\Gamma^\alpha\psi(y).
  \nonumber
\end{eqnarray}
After Fourier transform, in momentum space one obtains
\begin{eqnarray}
{\cal S}_{\rm NJL}&=&-\frac{N_cg^2}{4G}\int\frac{d^4q}{(2\pi)^4}
  \tilde M_\alpha^*(q)\tilde M^\alpha(q)+\int\frac{d^4p}{(2\pi)^4}
  \bar{\tilde\psi}(p)(\slp-m_q)\tilde\psi(p)\strut\\&&\strut
  -\frac{N_cg^2}{2G}\int\frac{d^4p}{(2\pi)^4}\int\frac{d^4p'}{(2\pi)^4}
  \bar{\tilde\psi}(p)\tilde\Delta\pfrac{p+p'}2\left(\tilde M_\alpha(p-p')
  +\tilde M_\alpha^*(p-p')\right)\Gamma^\alpha\tilde\psi(p').\nonumber
\end{eqnarray}
where the symbols with tilde are used for the Fourier transformed quantities.

\subsection{Mean field approximation}
Out of the many different contributions obtained after Fierz rearrangement,
the mean field approximation makes a choice that is phenomenologically
justified. We can expand the physical mesonic fields
$\tilde\sigma(p)=\bar\sigma+\delta\tilde\sigma(p)$ and
$\vec{\tilde\pi}(p)=\delta\vec{\tilde\pi}(p)$ about the vacuum expectation\
value $\bar\sigma=\langle\sigma\rangle$ where the expansion coefficient to
zeroth order is the mean field approximation. This approximation without any
information about possible correlations is sufficient for our means as it
leads directly to the mass gap equation. In this approximation one obtains
the simplified NJL action
\begin{eqnarray}
{\cal S}_{\rm NJL}&=&-\frac{N_cg^2\bar\sigma^2}{4G}V^{(4)}
  +\int\frac{d^4p}{(2\pi)^4}\bar{\tilde\psi}(p)(\slp-m_q)\tilde\psi(p)
  -\frac{N_cg^2\bar\sigma}{G}\int\frac{d^4p}{(2\pi)^4}\bar{\tilde\psi}(p)
  \tilde\Delta(p)\tilde\psi(p)\nonumber\\
  &=&-\frac{N_cg^2\bar\sigma^2}{4G}V^{(4)}
  +\int\frac{d^4p}{(2\pi)^4}\bar{\tilde\psi}(p)(\slp-M_q(p))\tilde\psi(p)
\end{eqnarray}
with $G=2\tilde\Delta(0)$ and the unit space-time volume $V^{(4)}$, where
\begin{equation}\label{muq1}
M_q(p)=m_q+\frac{N_cg^2}G\tilde\Delta(p)\bar\sigma
  =m_q+\frac{N_cg^2\tilde\Delta(p)}{2\tilde\Delta(0)}\bar\sigma
\end{equation}
is the dynamical mass of the quark. The bosonization procedure yields
\begin{equation}
\frac{{\cal S}_{\rm bos}}{V^{(4)}}=-\frac{N_cg^2\bar\sigma^2}{4G}
  -i\int\frac{d^4p}{(2\pi)^4}\ln\det(\slp-M_q(p)),
\end{equation}
where $\det$ denotes the direct product of a functional and an analytical
determinant, the former in the Fock space transition between space-time
points $x$ and $y$, the latter in the Dirac and flavor indices. On the other
hand, one has
\begin{equation}
\ln\det(\slp-M_q(p))=\tr\ln(\slp-M_q(p))
  =\frac124N_f\ln\left(p^2-M_q^2(p)\right).
\end{equation}
Taking the variation of the action ${\cal S}_{\rm bos}$ with respect to
$\bar\sigma$ yields the latter quantity. Accounting for the dependence of 
$M_q(p)$ on $\bar\sigma$, we have
\begin{equation}
0=-\frac{N_cg^2\bar\sigma}{2G}-2iN_f\int\frac{d^4p}{(2\pi)^4}
  \frac{2M_q(p)}{p^2-M_q^2(p)}\frac{N_cg^2}G\tilde\Delta(p)\ \Rightarrow\
  \bar\sigma=-8iN_f\int\frac{d^4p}{(2\pi)^4}
  \frac{\tilde\Delta(p)M_q(p)}{p^2-M_q^2(p)}.
\end{equation}
Finally, this result can be re-inserted to Eq.~(\ref{muq1}) to obtain the
dynamical mass equation
\begin{equation}
M_q(p)=m_q-4iN_fN_cg^2\frac{\tilde\Delta(p)}{\tilde\Delta(0)}
\int\frac{d^4p'}{(2\pi)^4}\frac{\tilde\Delta(p')M_q(p')}{p^{\prime2}
  -M_q^2(p')}.
\end{equation}
In this way, we have derived this equation directly from the QCD Lagrangian.
Following Ref.~\cite{Frasca:2021yuu} we do not consider the explicit
dependence of $M_0=M_q$ on the momentum. We use this choice to obtain a
qualitative picture of the dynamics, giving up the possibility of wavefunction
renormalisation and the possibility to estimate of the size of the error.
After that, performing the Wick rotation, we obtain
the mass gap equation in Euclidean space,
\begin{equation}\label{gapq}
M_0=m_q+C\alpha_s\int_0^{\Lambda^2}
  \frac{M_0\tilde\Delta(p')p^{\prime2}dp^{\prime2}}{p^{\prime2}+M_0^2},\qquad
  C=4N_fN_c.
\end{equation}

\section{Quark confinement\label{sec3}}

\begin{figure}\begin{center}
\epsfig{figure=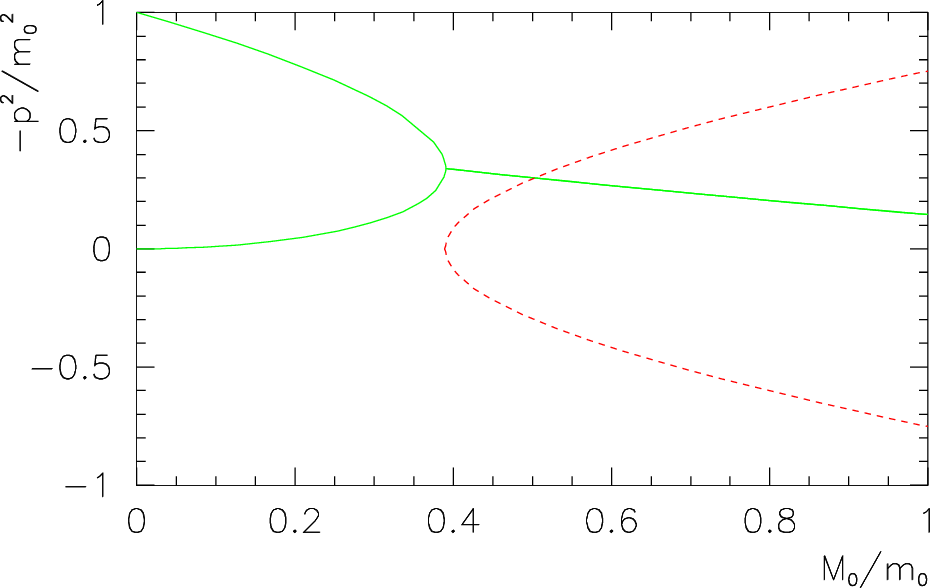, scale=0.8}
\caption{\label{lowestzero} \it The two lowest zeros of $-p^2=M_q(p)$ in
Euclidean domain in units of $m_0^2$, splitted into real parts (green straight
lines) and imaginary parts (red dashed lines), in dependence on the ratio
$M_0/m_0$. The zeros become complex for approximately $M_0/m_0>0.39$.}
\end{center}\end{figure}

The idea to understand quark confinement is strongly linked to the expected
behavior of the roots of the gap equation~(\ref{gapq}), i.e.\ the poles of the
quark propagator. The idea presented here is identical with the idea presented
in Ref.~\cite{Rezaeian:2004nf}, though without employing a general model for
the non-locality. To represent physical propagating degrees of freedom, for
these poles one should expect solutions on the real axis. The effect of the
gluonic interaction is to move such poles in the complex plane so that no
decay into such degrees of freedom is ever expected and the free quarks never
propagate. This moves the solution from a chiral condensate phase to a
confining phase for quarks at increasing coupling. Indeed, a full
comprehension of such roots can only be achieved through the non-approximated
gluon propagator $\tilde\Delta(p)$. Therefore, we performed an analysis of the
lowest zero of $\tilde\Delta(p)$, finding out that for $M_0<0.39m_0$ there are
two distinct real zeros while above these two zeros are given by two complex
conjugate numbers. This can be seen in Fig.~\ref{lowestzero}. The threshold at
$0.39 m_0$ should be seen as the point beyond which, mathematically speaking,
a chiral condensate could possibly appear. We observe that the result depends
critically on the mass gap value $m_0$. In turn, this value depends on an
arbitrary integration constant and, therefore, should be fixed by the
experiment. This situation is similar to that of the constant $\Lambda$
entering in asymptotic freedom, and it is possible that these two constants
are related. Our best choice to fix this value is via the mixed gluonic-quark
state f$_0$(500) that could in principle be identified with the $\sigma$ meson
in the NJL model, giving rise to the breaking of chiral symmetry.

The gap equation~(\ref{gapq}) can be simplified by choosing $m_q=0$, and the
factor $M_q$ can be cancelled from the numerator to remove the trivial case.
This technique permits to find the fixed point solution of the given iterative
integral equation avoiding the awkward issue to resum all the iterates. One
obtains
\begin{eqnarray}
1&=&C\alpha_s\int_0^{\Lambda^2}
  \frac{\Delta(p')p^{\prime2}dp^{\prime2}}{p^{\prime2}+M_0^2}
  \ =\ C\alpha\sum_{n=0}^\infty\int_0^{\Lambda^2}
  \frac{B_np^{\prime2}dp^{\prime2}}{(p^{\prime2}+(2n+1)^2m_0^2)
  (p^{\prime2}+M_0^2)}\nonumber\\
  &=&C\alpha_s\sum_{n=0}^\infty\int_0^{\Lambda^2}
  \frac{B_ndp^{\prime2}}{(2n+1)^2m_0^2-M_0^2}
  \left(\frac{(2n+1)^2m_0^2}{p^{\prime2}+(2n+1)^2m_0^2}
  -\frac{M_0^2}{p^{\prime2}+M_0^2}\right)\\
  &=&C\alpha_s\sum_{n=0}^\infty\frac{B_n}{(2n+1)^2m_0^2-M_0^2}
  \left((2n+1)^2m_0^2\ln\left(1+\frac{\Lambda^2}{(2n+1)^2m_0^2}\right)
  -M_0^2\ln\left(1+\frac{\Lambda^2}{M_0^2}\right)\right).\kern-6pt\nonumber
\end{eqnarray}
For $\Lambda=1\GeV$ and $m_0=0.512(15)\GeV$ one obtains $M_0=0.427(29)\GeV$ as
in Ref.~\cite{Frasca:2021yuu} which is clearly above the threshold
$0.39m_0=0.2\GeV$ and, therefore, indicates quark confinement.

\subsection{Understanding quark confinement}

As we have seen, with increasing coupling of the theory the quark confinement
arises at the point where the chiral condensates of different flavors perform
a transition to a confined phase with an instanton liquid of gluon degrees of
freedom. Neglecting the bare quark masses we have found the critical effective
quark mass for which such a transition happens. Our choice of the ground state
for the gluon field represents quite well a Fubini
instanton~\cite{Fubini:1976jm}. Therefore, we expect that the chiral
condensate changes into an instanton ground state that could condensate into a
liquid.

The presence of the instanton liquid removes single quarks from the physical
spectrum of the theory. From the mathematical point of view this means that
the poles in the quark propagator become complex. The vacuum of the theory
appears to undergo a series of phase transitions while the gluon sector
generates a mass gap by itself in a dynamical way. The presence of the mass
gap in the gluon sector is pivotal for the appearance of the phase transitions
in the quark sector and, in the last instance, to the confinement of quarks.

\section{Conclusions and Outlook\label{conc}}
There are different approaches to understand the confinement of quarks. One of
these is given by solutions of the gap equation of the dynamical quark mass.
With a reasonable UV cutoff of $\Lambda=1\GeV$ and the glueball spectrum
starting at the mass $m_0=0.512(15)\GeV$ of the $f_0(500)$ resonance, the gap
equation provides a value $M_0=0.427(29)\GeV$ for the dynamical quark mass
which turns out to be too large to allow for real valued poles of the quark
propagator. As a consequence, free quarks are no longer physical states of the
theory and the quarks can be expected to be confined. Even though the multiple
approximations applied in this approach do not allow for quantitative
estimates, we have shown how this scenario is realized in the low-energy
regime of QCD by taking into account the 't~Hooft limit. By doing so, the
low-energy limit of QCD turns out to be a well-defined non-local NJL model
with all the parameters obtained from QCD. This entails a scenario where
several condensates are formed that are expected to provide confinement in a
regime of very low momentum and strong coupling.

Having a low-energy limit of QCD permits to do several computations to be
compared with experiments. Indeed, our first application was to the $g-2$
problem~\cite{Frasca:2021yuu} with a very satisfactory agreement with data.
Further research has to show whether and to what extent our description of
quark confinement depends on the given parameter values and whether a stricter
derivations of observables allows for a comparison with experiments.

\section{Acknowledgements\label{Ack}}
The research was supported in part by the European Regional Development Fund
under Grant No.~TK133.

\end{document}